\documentclass
[aps,pra,reprint,showpacs,floatfix,amsmath,amssymb,10pt]{revtex4-1}

\usepackage{graphicx}
\usepackage{epstopdf}
\usepackage{dcolumn}
\usepackage{bm}
\usepackage{amsmath}

\def\gappeq{\mathrel{ \rlap{\raise.5ex\hbox{$>$}}
                      {\lower.5ex\hbox{$\sim$}}  } }
\def\lappeq{\mathrel{ \rlap{\raise.5ex\hbox{$<$}}
                      {\lower.5ex\hbox{$\sim$}}  } }
                      
\begin{document}
\title{Coherent tunneling via adiabatic passage in a three-well Bose-Hubbard system}
\date{\today}

\author{C. J. Bradly}
\affiliation{School of Physics, The University of Melbourne, Parkville,
3010, Australia}
\author{M. Rab}
\affiliation{School of Physics, The University of Melbourne, Parkville,
3010, Australia}
\author{A. D. Greentree}
\affiliation{School of Physics, The University of Melbourne, Parkville,
3010, Australia}
\author{A. M. Martin}
\affiliation{School of Physics, The University of Melbourne, Parkville,
3010, Australia}

\begin{abstract}
We apply the Bose-Hubbard Hamiltonian to a three-well system and show analytically that coherent transport via adiabatic passage (CTAP) of $N$ non-interacting particles across the chain is possible. We investigate the effect of detuning the middle well to recover CTAP when on-site interparticle interactions would otherwise disrupt the transport. The case of small interactions is restated using first-order perturbation theory to develop criteria for adibaticity that define the regime where CTAP is possible. Within this regime we investigate restricting the Hilbert space to the minimum necessary basis needed to demonstrate CTAP, which dramatically increases the number of particles that can be efficiently considered. Finally, we compare the results of the Bose-Hubbard model to a mean-field three-mode Gross-Pitaevskii analysis for the equivalent system.
\end{abstract}
\pacs{03.75.-b, 05.30.Jp}
\maketitle

\section{INTRODUCTION}
The Bose-Hubbard model provides a useful theoretical and experimental platform to study the properties of strongly interacting bosonic systems. In particular, considerable effort has been applied to the properties of interacting bosons in a lattice, especially in the context of ultracold atom research \cite{Greiner,Fisher}. The Bose-Hubbard model has also been extensively employed in the investigation of the properties of ultracold bosonic atoms confined in a double-well potential \cite{Smerzi,Milburn,Leggett,Albiez,Schumm,Gati,Folling}. Additionally, the work on double-well systems has been extended to triple well systems \cite{Mossmann06,Mossmann08,Streltsov,Viscondi11,Viscondi10}. Despite the relative simplicity of the Bose-Hubbard model it provides a rich space of phenomenology to explore, much of which is inaccessible to present analytical techniques due to the growth in size of the Hilbert space for a large number of particles. Hence there is a need to develop analytically tractable problems for Bose-Hubbard systems to gain insight into the analytically intractable problems.
Here we utilize the Bose-Hubbard model to investigate the adiabatic transport of ultracold bosonic atoms confined in a triple-well potential. 

For a single quantum particle in a three-well system it is possible to transport the particle between the left and right wells such that the probability of finding it at any time in the classically accessible state in the middle well is negligible.  The ideas underpinning such a transport mechanism stem from stimulated Raman adiabatic passage (STIRAP) \cite{Shore,Oreg,Kuklinski,Gaubatz,Bergmann}. In general, STIRAP is employed to transfer population between two atomic states, via an intermediate state, with the occupation of the intermediate state strongly suppressed, and is used in quantum optics for coherent internal state transfer \cite{Bergmann,Weitz,Wynar,Winkler}. In analogy with STIRAP, it is possible to adiabatically transport a quantum particle from the left well to the right well such that the occupation of the middle well is exponentially suppressed. For such a system the three states are typically the ground states of each of the three wells and the coupling between the states is the tunneling rates between the wells, although it is possible to transport between excited states of each well \cite{CTAP3}. This spatial transport is referred to as coherent tunneling via adiabatic passage (CTAP) and has been proposed to transport single atoms \cite{Eckert,Deasy}, Cooper pairs \cite{Siewert}, spin states \cite{Ohshima}, electrons \cite{Zhang,CTAP1,Fabian,CTAP2,CTAP3}, ultracold atoms in atom-chip waveguides \cite{O'Sullivan} and photons in three-channel optical waveguides \cite{Longhi_prop}. For the case of three-channel optical waveguides, CTAP for massless particles has been demonstrated \cite{Longhi,Della}.


\begin{figure}[b!]
\centering
\includegraphics[width=\columnwidth ]{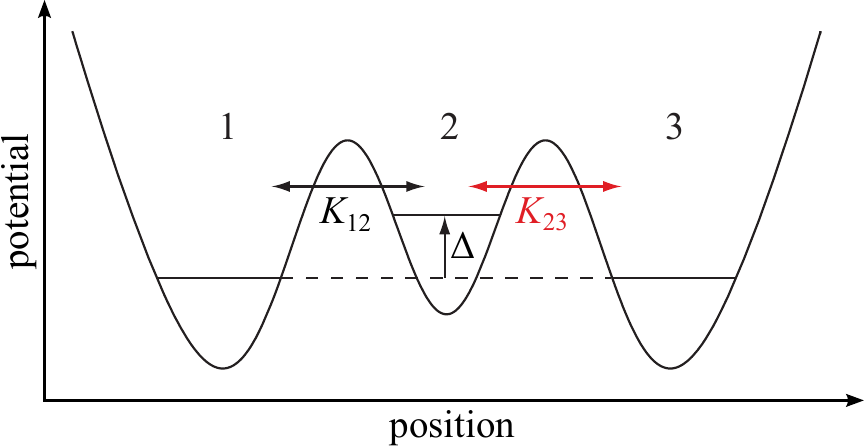}
\caption{
Three-well system where each well is assumed to have only one accessible energy level. Wells one and three are degenerate, whilst the middle-well ground state is shifted relative to the other two by a detuning, $\Delta$. The tunneling rate between wells $i$ and $j$ is $K_{ij}$.}
\vspace{-0.0cm}
\label{fig:System}
\end{figure}

\begin{figure*}[t!]
\centering
\includegraphics[width=\textwidth]{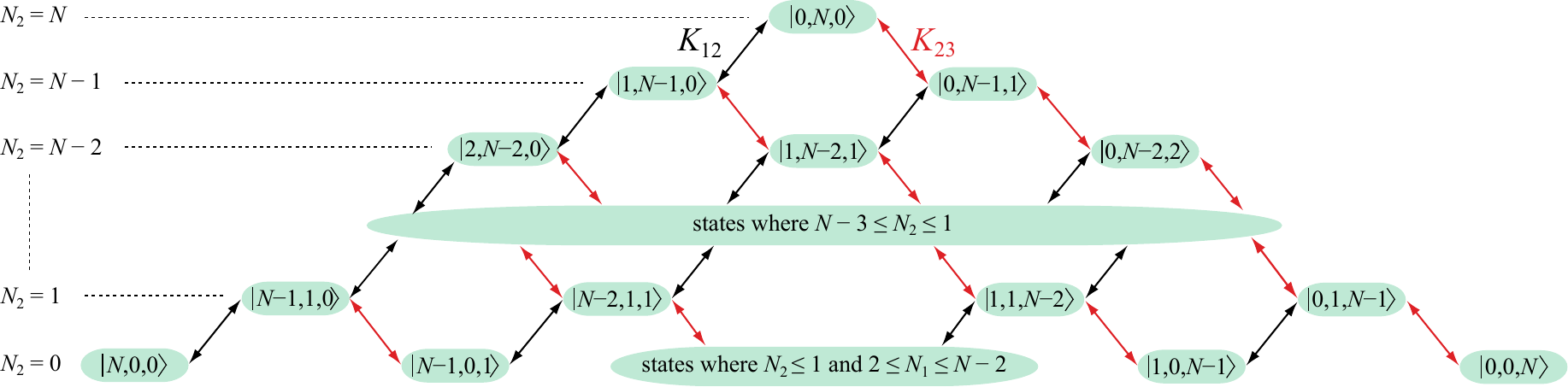}
\caption{
Hilbert space structure for $N$ particles in three wells. Each layer of the Hilbert space is denoted by $N_2$ and the coupling between layers is via $K_{12}$ (black arrows), which couples sites $1$ and $2$, and $K_{23}$ (red arrows) which couples sites $2$ and $3$.}
\vspace{-0.5cm}
\label{fig:HilbertSpace}
\end{figure*}

Recently, considerable attention has been paid to the CTAP of Bose-Einstein condensates (BECs) in three-well systems \cite{Graefe,Wang,Liu,Rab,Nesterenko}. These studies have primarily focused on a three-mode model of the system (schematically shown in Fig.~\ref{fig:System}), in both the linear and nonlinear regimes, to determine the robustness of CTAP. Additionally, numerical simulations, using the mean-field Gross-Pitaevskii equation, have been performed to model CTAP of a three-dimensional BEC containing $2000$ interacting atoms \cite{Rab}.  Here we take a different approach to studying adiabatic transport in many-particle systems by considering a three-site Bose-Hubbard model with $N$ interacting bosons. 
We find that the non-interacting limit of this system maps onto CTAP of a single particle along a chain of $2N+1$ sites via an alternating coupling protocol \cite{Petrosyan,Jong,Hill}. This mapping allows us to apply several analytical results to the problem, which would otherwise be computationally intensive, since the size of the Hilbert space basis is of order $N^2$.

This paper is organized as follows: In Section II the three-site Bose-Hubbard model is introduced and the non-interacting null state is found analytically. The breakdown of CTAP in the Bose-Hubbard model is then investigated numerically in the presence of interactions. In Section III, perturbative methods are used to demonstrate how detuning of the middle well can maintain CTAP once interactions are introduced into the system. Section IV numerically investigates the fidelity of CTAP as a function of the interactions and the detuning of the middle well. This is compared with an analytic determination of the regimes of high fidelity based on the analysis presented in Section III. Additionally, a comparison between the results obtained for the Bose-Hubbard model and three-mode non-linear mean-field calculations is made. Finally, Section V summarizes the key results obtained in this work and discusses possible avenues for further theoretical and experimental studies.

\section{THREE WELL BOSE-HUBBARD SYSTEM}

To begin the analysis we consider a three-well system, schematically shown in Fig.~\ref{fig:System}, subject to the Bose-Hubbard Hamiltonian
\begin{alignat}{1} 
\hat{H} =& - K_{12}\left( {\hat{a}_1}^\dagger \hat{a}_2 + {\hat{a}_2}^\dagger \hat{a}_1\right)-K_{23}\left( {\hat{a}_2}^\dagger \hat{a}_3 + {\hat{a}_3}^\dagger \hat{a}_2\right) \nonumber \\
+ & \frac{U}{2}\sum_{i=1}^3 \hat{n}_i (\hat{n}_i - 1) + \sum_{i=1}^3 \epsilon_{i} \hat{n}_i,
\label{eq:BHH}
\end{alignat}
where ${\hat{a}_i}^\dagger$, ${\hat{a}_i}$ and $\hat{n}_i$ are the bosonic creation, annihilation and number operators for site $i$, $K_{12}$ ($K_{23}$) is the tunneling matrix element between sites $1$ and $2$ ($2$ and $3$), $U$ is the particle-particle interaction energy and $\epsilon_{i}$ is the ground state energy of site $i$. We only consider tunneling between nearest neighbors so we set $K_{13}=0$. We define the energy scale of the system by the ground state energies of wells 1 and 3, i.e., $\epsilon_1=\epsilon_3=0$ but we include a (positive) detuning of well 2 so that $\epsilon_2=\Delta$. 

There are $M=(N+1)(N+2)/2$ ways to distribute $N$ particles amongst three wells and these configurations are the Fock basis states $|N_1,N_2,N_3\rangle$ for the Hilbert space of the system. The restriction on the individual well occupation numbers $N_i$ is $\sum_{i=1}^3 N_i =N$. Figure \ref{fig:HilbertSpace} shows the Hilbert space structure for the system, with each layer of the Hilbert space defined by $N_2$. The Fock basis states have a natural ordering based on the individual occupations of each well. This ordering gives rise to a triangular pattern, with layers defined by the number of particles in well 2, and the left to right ordering by the distribution of the particles in wells 1 and 3, as depicted in Fig.~\ref{fig:HilbertSpace}. In general, the system can take multiple, looped paths through the Hilbert space, giving rise to non-trivial dynamics \cite{JongGreentree}. However, nonzero detuning on well 2 makes higher states difficult to access by the system, reducing the contribution from these looped paths. The total adiabatic dynamics can, in the appropriate limit of small interactions, be very well described by the lowest two layers of the Hilbert space.

\subsection{Non-interacting case}
The Hamiltonian \eqref{eq:BHH} cannot, in general, be solved analytically and must be investigated numerically. However, before investigating CTAP of interacting bosons in a three-well system, we can gain significant insight into the problem by considering the non-interacting case, which can, in the adiabatic limit, be solved analytically.

In the absence of interactions, the system closely resembles a single particle in a chain of $2N+1$ sites with an alternating coupling scheme \cite{Petrosyan,Jong,Shore,Hill}. It is therefore possible to write down the null state of the Hamiltonian \eqref{eq:BHH} for a system of $N$ non-interacting particles in a three-well system:
\begin{equation} \label{eq:nullstate}
|{\cal D}_{0}\rangle_N = \frac{1}{{\cal N}} \sum_{k=0}^N (-1)^k \sqrt{\frac{N!}{k!(N-k)!}} \left(\frac{K_{23}}{K_{12}}\right)^k
|k,0,N-k\rangle ,
\end{equation}
with normalization
\begin{equation}
{\cal N} = \left[\sum_{k=0}^N \frac{N!}{k!(N-k)!} \left(\frac{K_{23}}{K_{12}}\right)^{2k}\right]^\frac{1}{2} = \left(1+\frac{K_{23}^2}{K_{12}^2}\right)^{\frac{N}{2}}.
\end{equation}
The null state is unique for $\Delta \not =0$. If the system is prepared in the $|{\cal D}_{0}\rangle_N$ eigenstate and evolved adiabatically, then the system will only occupy states in the $N_2=0$ layer of the Hilbert space (see Fig.~\ref{fig:HilbertSpace}), so there is vanishing probability of detecting a particle in well 2. The system's Hilbert space trajectory will also explore the $N_2=1$ layer, since these are necessary to couple to the $N_2=0$ states, but occupation of the $N_2=1$ states is zero in the adiabatic limit. Hence, under adiabatic evolution, $N$ particles initially in state $|N,0,0\rangle$ ($K_{12}=0$,  $K_{23}=1$) can be transported to the state $|0,0,N\rangle$ ($K_{12}=1$, $K_{23}=0$), via the states $|N-k,0,k\rangle$, with $0<k< N$. This transformation must be performed adiabatically to maintain the system in the null state, discussed further in Section III. The key issue here is that by changing the couplings appropriately we can transport the system through the Hilbert space. This is equivalent to transporting the particles from well 1 to well 3 without intermediate occupation of well 2.

\begin{figure}[t!]
\includegraphics[]{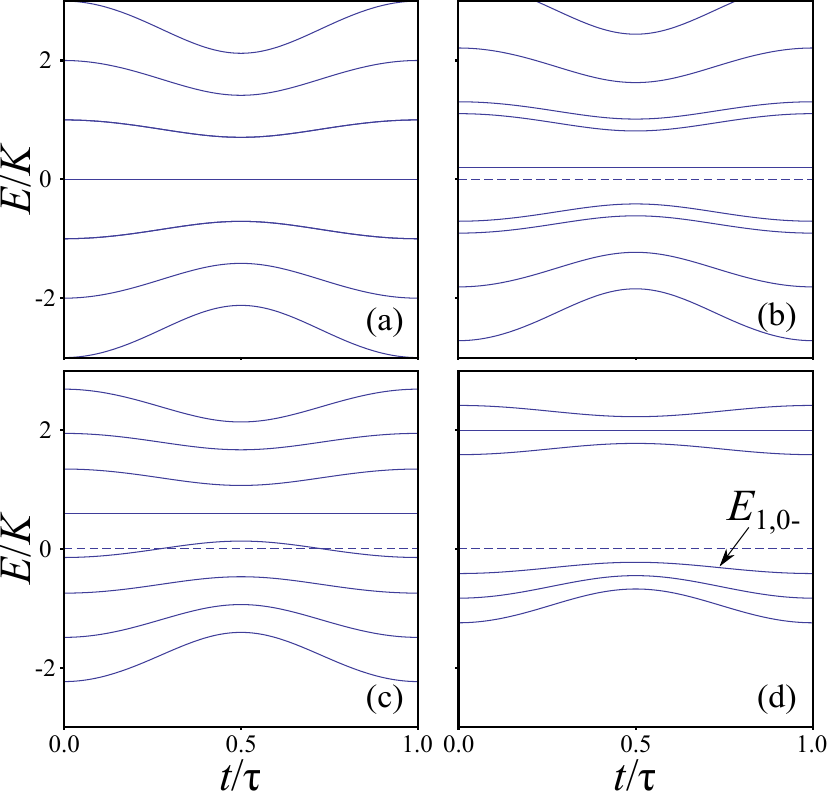}
\caption{Non-interacting energy eigenvalue spectrum for $N=3$ using the squared-sinusoidal pulsing protocol Eq.~\eqref{eq:pulses}. (a) With no detuning, $\Delta/K=0$, the null state (dashed) is degenerate. (b) For small detuning, $\Delta/K=0.2$, the states are lifted slightly, removing the degeneracy. (c) Moderate detuning, $\Delta/K=0.6$, lifts states enough that some pass through the null state and cross at certain times, potentially interrupting the adiabatic transport. (d) Large detuning, $\Delta/K=2$, pushes those crossing states past the null state at the cost of bringing the $E_{1,0-}$ state closer to the null state at $t=\tau/2$.}
\label{fig:Spectrum}
\end{figure}

In the non-interacting limit, the full spectrum of $M$ eigenenergies is
\begin{equation} \label{eq:eigenenergies}
E_{k,j\pm} = \frac{k}{2}\Delta \pm \frac{k-2j}{2}\sqrt{\Delta^2 + 4(K_{12}^2 + K_{23}^2)},
\end{equation}
for integers $k$ and $j$, where $0 \leq k \leq N$, $0\leq j \leq \lfloor k/2 \rfloor$ and $E_0 \equiv E_{0,0} =0$ denotes the null state energy. The corresponding eigenstates $|{\cal D}_{k,j\pm}\rangle_N$, hereafter denoted simply as $|{\cal D}_\alpha\rangle$, can be found in closed form for any $N$. The null state \eqref{eq:nullstate} exists for any detuning $\Delta$ but, as indicated by Eq.~\eqref{eq:eigenenergies}, it is not unique when $\Delta=0$, namely when $k$ is odd and $j=0$ there are eigenenergies that are integer multiples of $\Delta$ which vanish when $\Delta \to 0$. In this case there are $(N+1)/2$ null states for odd $N$, and $N/2$ null states for even $N$. These extra null states have contributions from basis states with $N_2>0$ so if the system evolved into one of them then the CTAP transport would be disrupted. In addition, the adiabatic coupling (discussed in Section III) between the preferred null state and these extra null states is undefined, indicating that these degenerate states are unavoidable. To prevent this we can introduce a small positive detuning to lift these degeneracies, as shown in Fig.~\ref{fig:Spectrum}(b). This will isolate the desired null state \eqref{eq:nullstate} and improve the fidelity of CTAP. However, larger detunings, shown in Figs.~\ref{fig:Spectrum}(c,d), can introduce new problems related to the adiabatic couplings of other eigenstates. We address this issue in Section III B. 

If well 2 is not detuned, i.e.~$\Delta=0$, then CTAP is still possible provided we initialize the system correctly and evolve the system adiabatically, despite the extra null states \cite{Pyragas}. This is because when $K_{12}=1$, $K_{23}=0$ or vice-versa, the extra null states are in a superposition of more than one basis state, whereas the preferred null state, Eq.~\eqref{eq:nullstate}, is the only eigenstate that is exactly a single basis state. For the case of two particles this is shown explicitly below. Hence, we conclude that these extra null states do not interfere with CTAP and demonstrate this for the case of two particles below \cite{Hill}.

\begin{figure}[t!]
\centering
\includegraphics[width=\columnwidth ]{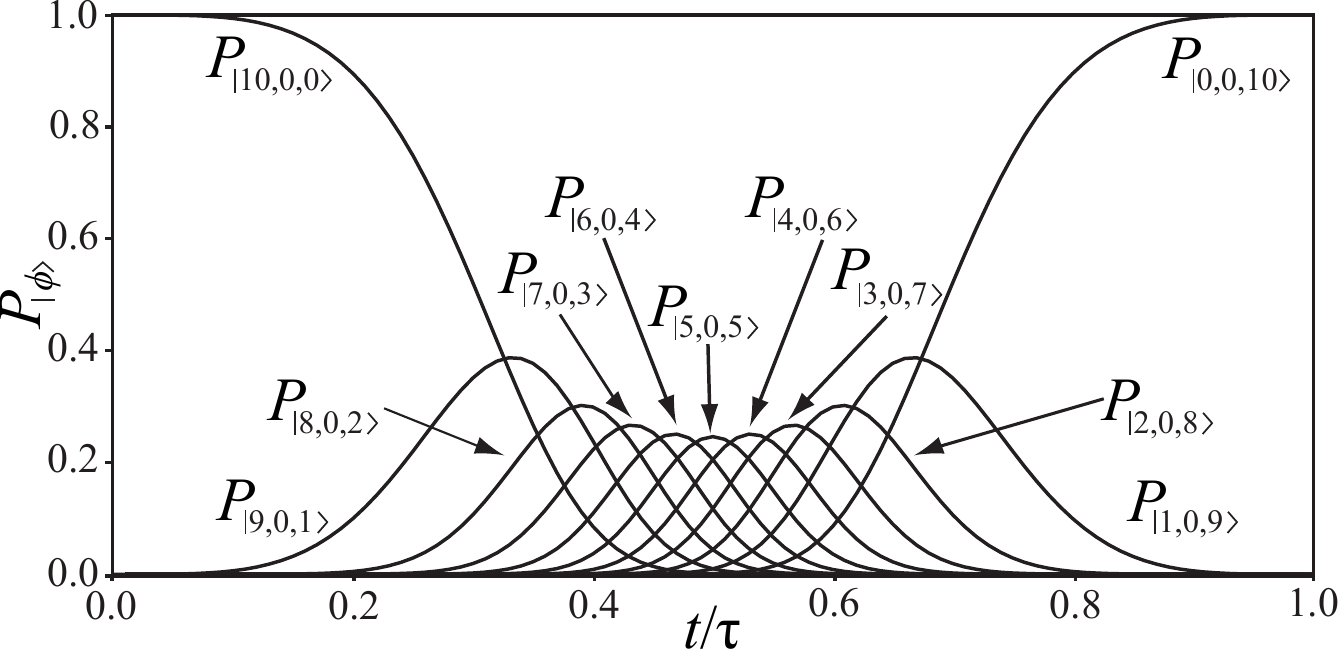}
\caption{Basis state occupation probabilities $P_{{|\phi \rangle}}(t)$ for the non-interacting null state for ten particles. The evolution is for the squared-sinusoidal pulsing, with $U=0$ and $\Delta=0$, in the adiabatic limit ($\tau$ large). The basis states with $N_2>0$ are negligibly occupied throughout the pulsing.}
\label{fig:10ParticleProbabilitiesU0}
\vspace{-0.5cm}
\end{figure}

The time-evolution of the system is determined by the pulsing of the potential barriers $K_{12}$ and $K_{23}$. If all the particles are initialized in well 1, i.e.~$\Psi(0)=|N,0,0\rangle$, then to induce the required population transfer we pulse the left and right barriers in counter-intuitive order. The precise form of the pulsing is not as important for CTAP as the order, so for convenience we choose squared-sinusoidal pulsing
\begin{equation} \label{eq:pulses}
K_{12}(t) = K \sin^2 \left(\frac{\pi t}{2\tau}\right),\quad 
K_{23}(t) = K \cos^2\left(\frac{\pi t}{2\tau}\right),
\end{equation}
where $K$ is the maximum tunneling rate and $\tau$ is the total pulse time. From Eq.~\eqref{eq:nullstate}, at $t=0$ ($t=\tau$), the probability of being in the $|N,0,0\rangle$ ($|0,0,N\rangle$) state is $1$. The probability that the system is in a particular basis state ${|\phi \rangle}=| N_1,N_2,N_3\rangle$ as a function of time is $P_{{|\phi \rangle}} (t)=|\langle N_1,N_2,N_3|\Psi(t) \rangle|^2$. These probabilities describe the trajectory the system takes through the Hilbert space over the course of the pulsing. 
Fig.~\ref{fig:10ParticleProbabilitiesU0} shows the population transfer, for ten particles, from the $|10,0,0\rangle$ state to the $|0,0,10\rangle$ state via the null state, Eq.~\eqref{eq:nullstate}, in the adiabatic limit.


To demonstrate CTAP in the non-interacting case we revisit the case of one particle \cite{Eckert,Deasy,Zhang,CTAP1,Fabian}. The single particle basis states are $|0,1,0\rangle$, $|1,0,0\rangle$, $|0,0,1\rangle$ for which the Hamiltonian \eqref{eq:BHH} is expressed as the matrix
\begin{equation} 
\hat{H}_1 =
\begin{pmatrix}
 \Delta 	& -K_{12}   & -K_{23}  	\\
 -K_{12} 	& 0  		& 0 		\\
 -K_{23}	& 0   		& 0
\end{pmatrix}
,
\label{eq:BHH1}
\end{equation}
and the null state, Eq.~\eqref{eq:nullstate}, is 
\begin{equation} \label{eq:nullstate1}
|{\cal D}_{0}\rangle = \frac{K_{23} |1,0,0\rangle-K_{12}|0,0,1\rangle}{\sqrt{K_{12}^2+K^2_{23}}},
\end{equation}
which has no contribution from $|0,1,0\rangle$. This is precisely the form of the canonical null state used for CTAP \cite{CTAP1} and is equivalent to the more familiar dark state in doubly driven three-level atoms \cite{Arimondo}.
Similarly, we may consider the two-particle system, which is spanned by the basis states $|0,2,0\rangle$, $|1,1,0\rangle$, $|0,1,1\rangle$, $|2,0,0\rangle$, $|1,0,1\rangle$ and $|0,0,2\rangle$. The Hamiltonian \eqref{eq:BHH} is 
\begin{widetext}
\begin{equation} 
\hat{H}_2 =
\begin{pmatrix}
	2 \Delta 		& -\sqrt{2}K_{12} 	& -\sqrt{2}K_{23} 	& 0 				& 0 		& 0 				\\
	-\sqrt{2}K_{12} & \Delta  			& 0  				& -\sqrt{2}K_{12}	& -K_{23} 	& 0 				\\
	-\sqrt{2}K_{23}	& 0 				& \Delta 			& 0 				& -K_{12}	& -\sqrt{2}K_{23} 	\\
	0 				& -\sqrt{2}K_{12}   & 0 				& 0 				& 0 		& 0 				\\
	0 				& -K_{23}  			& -K_{12}  			& 0 				& 0 		& 0					\\
	0				& 0 				& -\sqrt{2}K_{23} 	& 0 				& 0			& 0 
\end{pmatrix}
,
\label{eq:BHH2}
\end{equation}
\end{widetext}
and the null state, Eq.~\eqref{eq:nullstate}, becomes
\begin{equation} \label{eq:nullstate2}
|{\cal D}_{0}\rangle = \frac{K_{23}^2 |2,0,0\rangle - \sqrt{2}K_{12}K_{23}|1,0,1\rangle + K_{12}^2|0,0,2\rangle}{K^2_{12} + K^2_{23}}.
\end{equation}
This state also has no contribution from states with occupation of well 2 ($N_2>0$). As mentioned above, however, if $\Delta=0$ there is another null state
\begin{alignat}{1}
|{\cal D}_{0}'\rangle = & -\frac{1}{\sqrt{2}}|0,2,0\rangle + \frac{K_{12}^2}{\sqrt{2}(K^2_{12} + K^2_{23})} |2,0,0\rangle \nonumber \\ & + \frac{K_{12} K_{23}}{K^2_{12} + K^2_{23}}|1,0,1\rangle + \frac{K_{23}^2}{\sqrt{2}(K^2_{12} + K^2_{23})}|0,0,2\rangle.
\label{eq:extranullstate2}
\end{alignat}
If the system is in the primary null state, Eq.~\eqref{eq:nullstate2}, then it can be transformed from the state $|2,0,0\rangle$ when $K_{12}=0$ and $K_{23}=1$ to the state $|0,0,2\rangle$ when $K_{12}=1$ and $K_{23}=0$, but for both of these cases we see from Eq.~\eqref{eq:extranullstate2} that the secondary null state $|{\cal D}_{0}'\rangle$ always has a contribution from the $|0,2,0\rangle$ state. If both particles are initialized in well 2 then the system will be in the preferred null state $|{\cal D}_{0}\rangle$ and will remain in that state if evolved adiabatically.

\subsection{Interacting system}
The value of the Bose-Hubbard model is that it can model the interesting physics of real interacting atomic systems. The Hamiltonian for the three-well system Eq.~\eqref{eq:BHH} does not have a general null state for $U \not = 0$, so to find the behavior of the particles we must numerically integrate the time-dependent Schr\"odinger equation for the wavefunction $\Psi(t)$ of the system. Initializing the system as $\Psi(t=0)=|N,0,0\rangle$, we apply the CTAP pulsing sequence, Eq.~\eqref{eq:pulses}, and determine the probability $P_{|\phi \rangle}(t)$ that the system is in a particular basis state ${|\phi \rangle}$ as a function of time. If $\Psi(\tau)=|0,0,N\rangle$, then all the particles are said to have been transported across the chain, but for efficient CTAP we also require that throughout the pulsing protocol the contribution to the wavefunction from basis states with $N_2>0$ is minimized.

\begin{figure}[t!]
\centering
\includegraphics[width=\columnwidth ]{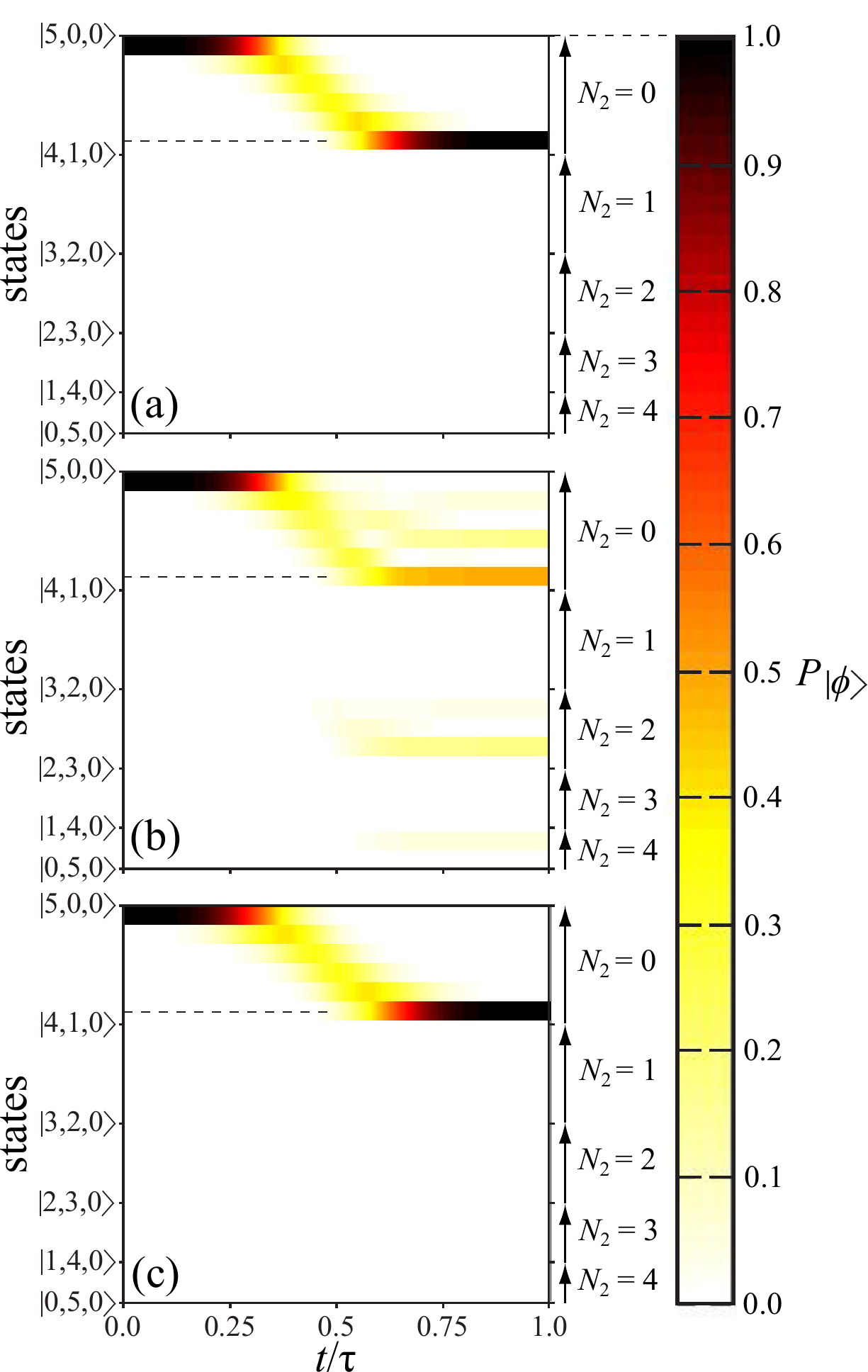}
\caption{Hilbert space trajectories $P_{{|\phi \rangle}}(t)$ for five-particles, using the pulsing sequence defined in Eq.~\eqref{eq:pulses}, with $\tau=500/K$. (a) $U=0$, $\Delta=0$, the system evolves as expected, remaining in the $N_2=0$ subspace. (b) $N U/K=0.03$, $\Delta=0$, strong interactions push the system out of the $N_2=0$ subspace, and reduce the fidelity of the transport even within the $N_2=0$ subspace. (c) $N U/K=0.03$, $\Delta/K=3$, a large detuning of the middle well restores CTAP and the particles are transported across the well with negligible occupation of the middle well. 
In each plot the states are presented along the $y$-axis and are ordered from bottom to top: $|0,5,0\rangle$; $|0,4,1\rangle$; $|1,4,0\rangle$; $|0,3,2\rangle$; $|1,3,1\rangle$; $|2,3,0\rangle$; $|0,2,3\rangle$; $|1,2,2\rangle$; $|2,2,1\rangle$; $|3,2,0\rangle$; $|0,1,4\rangle$; $|1,1,3\rangle$;$|2,1,2\rangle$; $|3,1,1\rangle$; $|4,1,0\rangle$; $|0,0,5\rangle$; $|1,0,4\rangle$; $|2,0,3\rangle$; $|3,0,2\rangle$; $|4,0,1\rangle$ and $|5,0,0\rangle$. The dashed line marks the desired final state $|0,0,5\rangle$. }
\label{fig:5ParticleHilbertTrajectories}
\vspace{-0.5cm}
\end{figure}

In Fig.~\ref{fig:5ParticleHilbertTrajectories} we plot $P_{{|\phi \rangle}}(t)$ for $\tau=500K^{-1}$, $N=5$ and various values of $U$ and $\Delta$. While five particles is still in the few-body regime, rather than the many-body regime, it is enough to demonstrate the interesting physics of the model while still being numerically tractable with standard techniques. Fig.~\ref{fig:5ParticleHilbertTrajectories}(a) shows transport for $U=0$, $\Delta=0$ and the system is confined to the lowest ($N_2=0$) layer of the Hilbert space, as expected from the analysis presented in the previous section. In Fig.~\ref{fig:5ParticleHilbertTrajectories}(b) the interactions are increased to $NU/K=0.03$, with $\Delta=0$, and there is a significant departure from the idealized null state transport observed in Fig.~\ref{fig:5ParticleHilbertTrajectories}(a). Specifically, the transport is no longer confined to the lowest layer of the Hilbert space and  $P_{|0,0,5\rangle}(\tau)<1$. However, it is possible to recover CTAP via the introduction of detuning, $\Delta$. For example, Fig.~\ref{fig:5ParticleHilbertTrajectories}(c), where $N U/K=0.03$ and $\Delta/K=3$, shows full population transfer between wells 1 and 3, such that $P_{|0,0,5\rangle}(\tau)\approx 1$. However, there is a limit to how effective this control is since increasing $\Delta$ reduces adiabaticity for constant pulsing time $\tau$, as discussed below. Increasing the pulsing time $\tau$ (short of the true adiabatic limit $\tau \to \infty$) does not restore CTAP when $\Delta=0$ for an interacting system and can even make it worse \cite{JongGreentree}. The non-adiabaticity in this regime is instead due to the CTAP state crossing with other eigenstates, similar to Fig.~\ref{fig:Spectrum}(c) for the non-interacting case. 


\section{PERTURBATION ANALYSIS}
\subsection{Large detuning limit}
We have seen that making $\Delta$ large relative to $NU$ restores CTAP for finite pulsing periods, motivating an analytic perturbation investigation of the Hamiltonian. As such, it is beneficial to rewrite the Bose-Hubbard Hamiltonian \eqref{eq:BHH} as $\tilde{H}={\hat H}/\Delta = {\hat H}^{(0)} + \tilde{U}{\hat V}$, where 
\begin{equation}
{\hat H}^{(0)} = - {\tilde K}_{12}\left( {\hat{a}_1}^\dagger \hat{a}_2 + {\hat{a}_2}^\dagger \hat{a}_1\right) - {\tilde K}_{23}\left( {\hat{a}_2}^\dagger \hat{a}_3 + {\hat{a}_3}^\dagger \hat{a}_2\right) + {\hat n}_2,
\end{equation}
\begin{equation}
{\hat V} = \frac{1}{2}\sum_{i=1}^3 {\hat n}_i ({\hat n}_i - 1),
\end{equation}
${\tilde K}_{12}=K_{12}/\Delta$, ${\tilde K}_{23}=K_{23}/\Delta$ and ${\tilde U}=U/\Delta$ is the small perturbation parameter if $N U/\Delta \ll 1$. In this Section the superscript $(0)$ denotes non-interacting quantities. It is important to note that although the effect of interactions changes with time as the distribution of particles in the system changes, ${\hat V}$ is time-independent in the basis of Fock states. 

In this limit it is possible to apply standard perturbation theory to calculate the effect of small interactions on the null state of the non-interacting case, Eq.~\eqref{eq:nullstate}. To first order, the interacting null state energy is 
\begin{equation} \label{eq:pertenergy}
E_0 = E_0^{(0)} +  \langle {\cal D}_{0}^{(0)}| {\hat V} | {\cal D}_{0}^{(0)}\rangle, 
\end{equation}
and the first-order correction to the null state wavefunction is
\begin{equation} \label{eq:pertnullstate}
|{\cal D}_0\rangle = |{\cal D}_{0}^{(0)}\rangle + \sum_{\alpha \neq 0}\frac{\langle {\cal D}_\alpha^{(0)}|{\hat V} |{\cal D}_{0}^{(0)}\rangle}{E_0^{(0)}-E_\alpha^{(0)}}|{\cal D}_\alpha^{(0)}\rangle,
\end{equation}
where $\alpha$ represents the $k,j\pm$ of Eq.~\eqref{eq:eigenenergies}. The eigenstates $|{\cal D}_\alpha^{(0)}\rangle$ of the non-interacting Hamiltonian corresponding to the energies of Eq.~\eqref{eq:eigenenergies} can be found in closed form, but for large $N$ it becomes more efficient to evaluate Eq.~\eqref{eq:pertnullstate} numerically. From these first order perturbative eigenstates it is then possible to calculate the basis state occupation probabilities $P_{|\phi\rangle}(t)$, to investigate the stability of CTAP in the presence of weak interactions. In the regime of small interactions relative to the detuning of well 2 this approach is more efficient than integrating the Schr\"odinger equation.

\begin{figure}
\centering
\includegraphics[width=\columnwidth ]{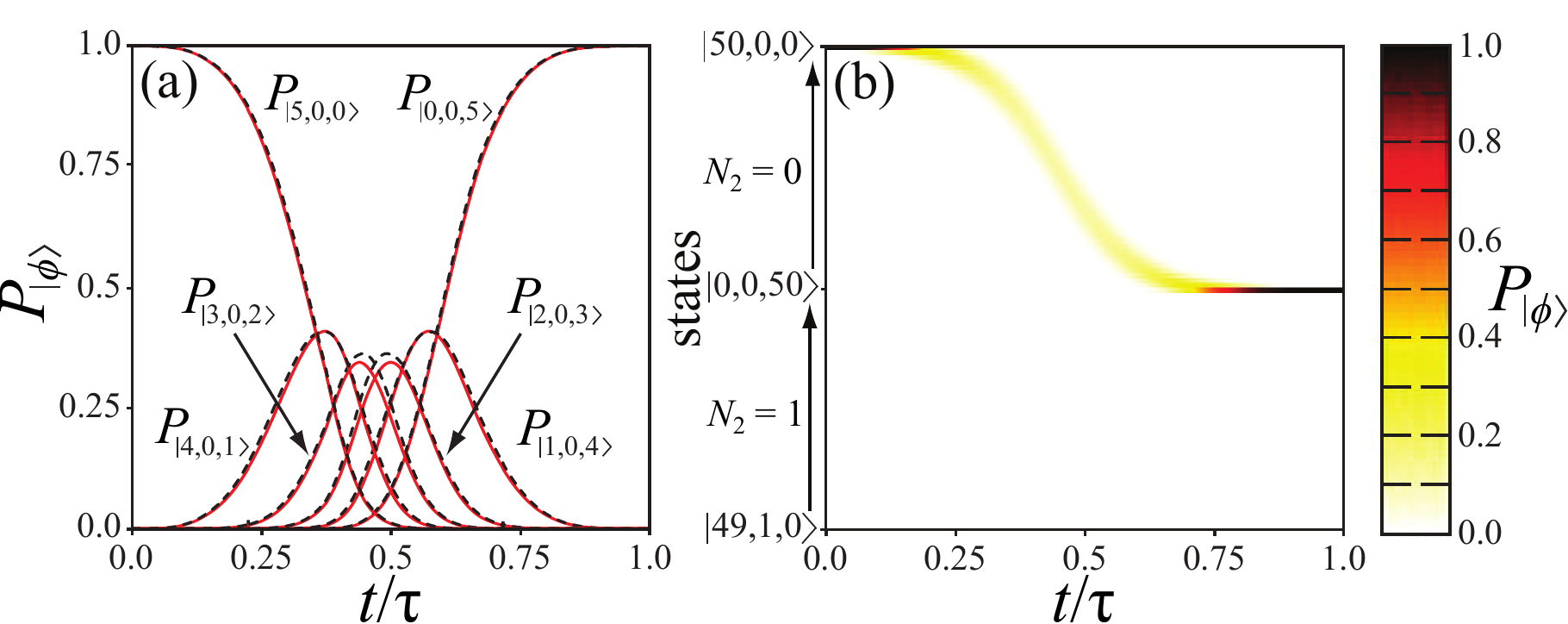}
\caption{(a) $P_{|\phi\rangle}(t)$ numerically calculated for the complete Hilbert space (red solid curve) and via perturbation theory (dashed black curve) for $N=5$, $\Delta/ K=1$, $\tau K =500$ and $NU/\Delta=0.01$. In both cases the occupation of states with $N_2>0$ is too small to be visible. (b) Hilbert space trajectory for a two-layer Hilbert space for $N=50$, $\Delta/K=1$, $\tau K=500$ and $NU/\Delta=0.01$.}
\label{fig:PerturbationAndRestricted}
\vspace{-0.5cm}
\end{figure}

Fig.~\ref{fig:PerturbationAndRestricted}(a) compares the perturbation theory method to a complete Hilbert space calculation for five particles. For small interactions relative to the detuning, transport from the state $|5,0,0\rangle$ to $|0,0,5\rangle$ is dominated by the $N_2=0$ states, just as in the non-interacting case. 

\subsection{Adiabaticity}

To determine when it is appropriate to describe transport of the particles from well $1$ to well $3$ via CTAP, the adiabaticity of the evolution needs to be verified. Increasing the pulsing time $\tau$ will in general make the transport more adiabatic, assuming no eigenstate crossings, but otherwise the adiabaticity of the transport within the null state $|{\cal D}_0\rangle$ depends on the coupling to the other energy eigenstates $|{\cal D}_\alpha\rangle$ and their energies $E_\alpha$. To quantify the adiabaticity of the transport we introduce the parameter  
\begin{equation} \label{eq:adiabaticA}
{\cal A}_{\alpha} = \frac{|\langle {\cal D}_\alpha| \dot{{\cal D}}_{0} \rangle|}{|E_\alpha-E_{0}|},
\end{equation}
and say that the evolution is adiabatic if ${\cal A}_{\alpha}\ll 1$ for all {$\alpha \not = 0$}. Eq.~\eqref{eq:adiabaticA} may be generalized to any pair of eigenstates but here we are interested in the coupling of the CTAP null state to other nearby eigenstates. It is also clear from Eq.~\eqref{eq:adiabaticA} that removing degeneracies via non-zero detuning $\Delta$ improves the adiabaticity of the transport. In particular, we focus on the large detuning limit $\Delta/K \gg 1$. 

In the absence of interactions, the eigenenergy that is closest to $E_0$ (for any $N$) is $E_{1,0-}^{(0)} = (\Delta - \sqrt{4 (K_{12}^2 + K_{23}^2) + \Delta^2})/2$, from Eq.~\eqref{eq:eigenenergies}. It is then possible to determine $|{\cal D}_{1,0-}^{(0)}\rangle$ and calculate the adiabatic coupling  ${\cal A}_{1,0-}$ for the squared-sinusoidal pulsing. Requiring ${\cal A}_{1,0-}\ll 1$ will suppress the coupling between this eigenstate and the null state. Since this is the eigenstate closest to the null state, ${\cal A}_{\alpha}$ will most likely be less than ${\cal A}_{1,0-}$ for all other eigenstates $\alpha$. From Fig.~\ref{fig:Spectrum}(d) we can also see that $E_{1,0-}^{(0)}$, and therefore ${\cal A}_{1,0-}$, is maximized at $t=\tau/2$ for the pulsing scheme given by Eq.~\eqref{eq:pulses}. 
For $\Delta/K \gg 1$ at $t=\tau/2$ 
\begin{equation} \label{eq:largeDnonint}
A_{1,0-} = \frac{2\pi\sqrt{N}\Delta}{\tau K^2} + \frac{\pi\sqrt{N}}{2\tau \Delta} + \frac{\pi\sqrt{N}K^2}{16\tau \Delta^3} + O(\Delta^{-5}).
\end{equation}
The leading order of Eq.~\eqref{eq:largeDnonint} is proportional to $\Delta$, so there is a limit on the detuning before the adiabatic coupling between the null state and the close state $|{\cal D}_{1,0-}^{(0)}\rangle$ becomes too large and the transport becomes non-adiabatic. To first order in $\Delta$, the requirement $A_{1,0-} \ll 1$ therefore becomes $\tau K/(2\pi \sqrt{N}) \gg \Delta/K$ for the non-interacting system. Note that in the true adiabatic limit, $\tau \to \infty$, the adiabaticity parameters ${\cal A}_\alpha$ will always vanish.

In the limit of small interactions relative to the detuning $\Delta \gg N U$, we again use perturbation theory to approximate the adiabaticity parameter by finding the first-order correction to all the eigenstates. Inserting Eqs.~\eqref{eq:pertenergy} and \eqref{eq:pertnullstate} in Eq.~\eqref{eq:adiabaticA} we obtain
\begin{eqnarray}
{\cal A}_\alpha & = & \left| \frac{\langle {\cal D}_{\alpha}^{(0)} | {\dot {\cal D}}_{0}^{(0)}\rangle}{E_{\alpha}^{(0)} - E_{0}^{(0)}} +  \frac{\langle {\cal{D}}_{\alpha}^{(1)}|{{\dot{\cal{D}}}}_0^{(0)} \rangle + \langle {\cal{D}}_{\alpha}^{(0)}|{\dot{\cal{D}}}_{0}^{(1)} \rangle}{E_{\alpha}^{(0)} - E_0^{(0)}}   \nonumber \right. \\
& & - \left. \langle {\cal{D}}_{\alpha}^{(0)}|{{\dot{\cal{D}}}}_{0}^{(0)} \rangle \left(E_{\alpha}^{(1)} - E_0^{(1)}\right) 
+ O  \left[ \left( \frac{N U}{\Delta} \right)^2 \right] \right|, \nonumber  \\
\label{eq:adiabatic_int}
\end{eqnarray}
where
\begin{alignat}{1}
|{\cal{D}}_{\mu}^{(1)}\rangle =&\sum_{\beta \neq \mu}\frac{\langle {\cal{D}}_\beta^{(0)}|{\hat V} |{\cal{D}}_{\mu}^{(0)}\rangle}{E_{\mu}^{(0)}-E_{\beta}^{(0)}}|{\cal{D}}_{\beta}^{(0)}\rangle, \\ 
\quad  E_{\mu}^{(1)}=&\langle {\cal{D}}_{\mu}^{(0)}| {\hat V} | {\cal{D}}_{\mu}^{(0)}\rangle,
\end{alignat}
which is valid for any pulsing protocol. As with the non-interacting case, the adiabaticity of the transport in the interacting case can be improved by increasing the pulsing time $\tau$, but for finite time, as required in experiment, Eq.~\eqref{eq:adiabatic_int}, can be used to show where the CTAP protocol becomes non-adiabatic and is used in the next section in conjunction with numerical results.

\section{RESTRICTED HILBERT SPACE}


For the case of small interactions relative to the detuning, if the system is evolved adiabatically with all particles initially in the state $|N,0,0\rangle$, the CTAP protocol only probes those states which have zero ($N_2=0$) or singular occupation ($N_2=1$) of well 2, i.e. the bottom two layers of the Hilbert space structure, shown in Fig.~\ref{fig:HilbertSpace}. If the system is not evolved adiabatically, or the interactions are large, then the full Hilbert space contributes to the dynamical evolution of the system, as seen in Fig.~\ref{fig:5ParticleHilbertTrajectories}(b). From this observation it is possible to consider the following question. If the Hilbert space is restricted to the lower layers ($N_2=0,1,\dots$), does the solution in this limit represent a valid approximation for the adiabatic transport between states $|N,0,0\rangle$ and $|0,0,N\rangle$, via CTAP? Hilbert space restrictions, if valid, will facilitate the investigation of systems for large $N$.

\begin{figure}
\centering
\includegraphics[width=\columnwidth ]{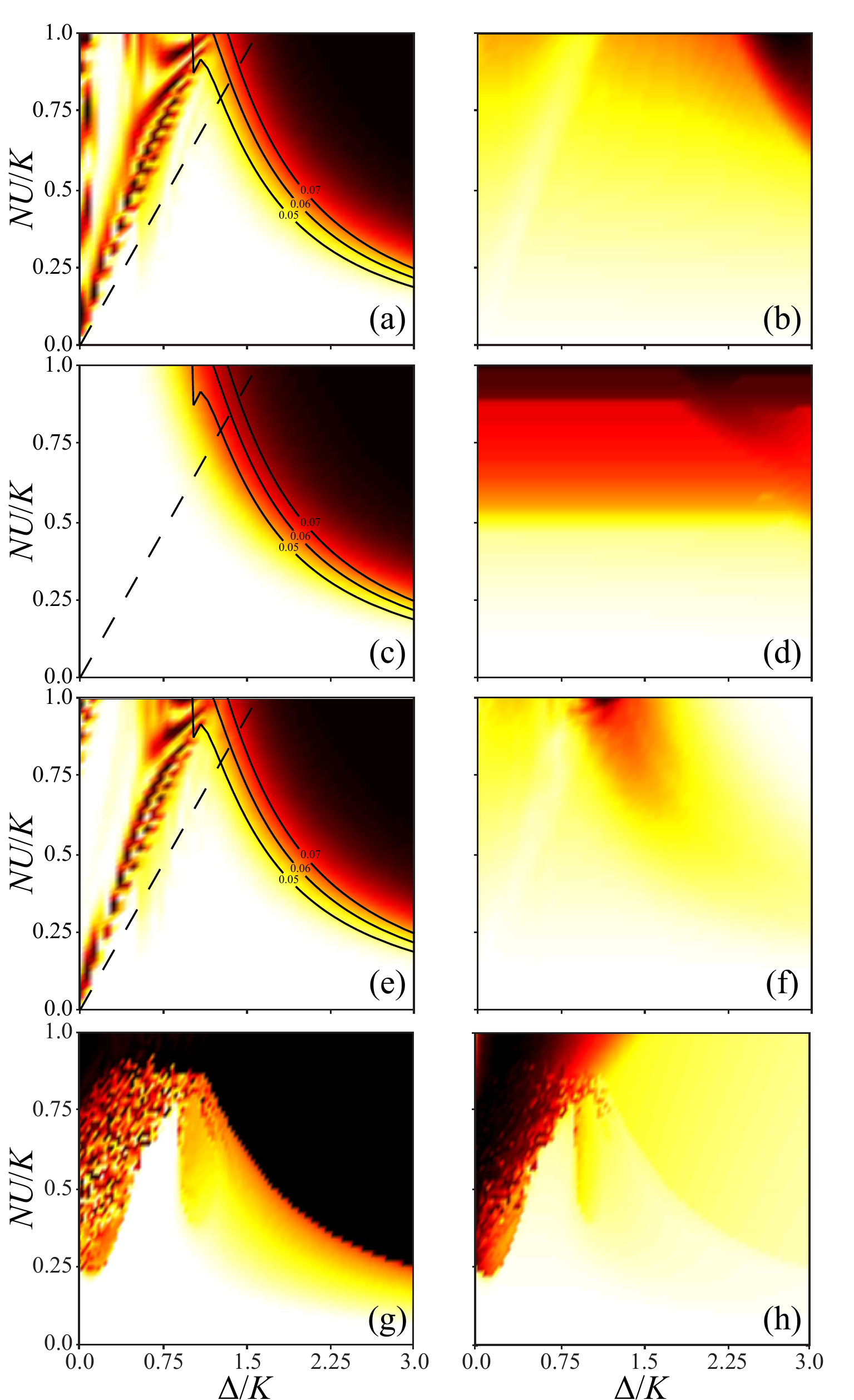}
\caption{(a,c,e) [white=$1$, black=$0$] $P_{|0,0,N\rangle}(t=\tau)$ for: (a) a full Hilbert space calculation; (c) a restricted two-layer ($N_2=0,1$) Hilbert space calculation and (e) a restricted four-layer ($N_2=0,1,2,3$) Hilbert space calculation. (b,d,f) [white=$0$] $\max P_{|N_1,N_2=1,N_3\rangle}(t)$ for: (b) a full Hilbert space calculation [black=$0.38$], with $N_2 = 1$; (d) a restricted two-layer ($N_2=0,1$) Hilbert space calculation [black=$0.35$], with $N_2=1$ and (f) a restricted four-layer ($N_2=0,1,2,3$) Hilbert space calculation [black=$0.45$], with $N_2=1$. In (a,c,e) the solid dashed line is given by the energy level criterion, Eq.~\eqref{eq:Delta_odd} and solid curves are for ${\cal A}_{1,0-}=0.05, 0.06, 0.07$ as calculated from Eq.~\eqref{eq:adiabatic_int}. (g) [white=$1$, black=$0$] Meanfield calculation of the final population of well three: $N_3(t=\tau)$.  (h) [white=0, black=1] Meanfield calculation of the maximum population of well two: $\max N_2(t)/N$. For the above plots $\tau K=500$ and for (a-f) $N=5$.}
\label{fig:5ParticleParameterSpace}
\vspace{-0.5cm}
\end{figure}

Rather than look at complete Hilbert space trajectories for every different value of $\Delta$ and $U$ to judge the validity of a restriction, we can consider two criteria to identify successful CTAP. Firstly, to determine if all the particles have been transported across the chain we calculate the probability that at the end of the pulsing $P_{|0,0,N\rangle}(\tau)\approx 1$. Secondly, to ensure that occupation of well 2 has been suppressed during the protocol we calculate the maximum occupation of a state with $N_2=1$, $\max P_{|N_1,N_2=1,N_3\rangle}(t)$. The latter can be thought of as a lower bound on occupation of well 2, since, if occupation of a $N_2=1$ state becomes too significant at any point, then the system may access higher basis states, which reduces the fidelity of the CTAP protocol. These two conditions are not mutually exclusive but in tandem they can be used to determine the fidelity of the population transfer. Figs.~\ref{fig:5ParticleParameterSpace}(c-f) plots these two quantities for a two- and four-layer Hilbert space restriction, compared to the full Hilbert space calculation, Fig.~\ref{fig:5ParticleParameterSpace}(a,b) for five particles.

The strongest restriction is to consider only the two bottom layers of the Hilbert space structure, states with $N_2=0,1$. There are only $2N+1$ basis states so the problem scales linearly with $N$, rather than with $N^2$ for the full system, and the Hamiltonian can be written in a tridiagonal form. The restricted Hamiltonian is still not invertible analytically for general $U$ but the Schr\"odinger equation can easily be integrated numerically for a system of several hundreds of particles, since the problem is then linear in $N$ (there are $2N+1$ basis states in the two bottom layers, of the Hilbert space structure). Furthermore, the null state of this restricted Hamiltonian in the non-interacting limit is exactly the null state for CTAP of a single particle along a chain of $2N+1$ sites \cite{Petrosyan,Jong}. Fig.~\ref{fig:PerturbationAndRestricted}(b) shows the Hilbert space trajectory for $N=50$ and $N U/\Delta=0.01$ using the two-layer restriction ($N_2=0$ and $N_2=1$). In this figure the only states with significant occupation are those with $N_2=0$, with essentially zero probability of the states with $N_2=1$ being occupied. Since only the states with $N_2=1$ couple to the rest of the Hilbert space structure (see Fig.~\ref{fig:HilbertSpace}), it could be expected that in this case a two layer Hilbert space is a good representation of the transport that would be observed in a full Hilbert space calculation. 

The two-layer restriction, while powerful, is only valid in certain regimes. In particular, Fig.~\ref{fig:5ParticleParameterSpace}(c) shows that the two-layer restriction  allows full population transfer in the case of high interaction strength but low detuning, while Fig.~\ref{fig:5ParticleParameterSpace}(d) shows that at higher interaction strengths well 2 is being occupied, regardless of detuning, in contrast to the full Hilbert space calculation, Figs.~\ref{fig:5ParticleParameterSpace}(a,b). Away from the small interaction limit, these results show that the two-layer restriction does not model CTAP well.

To incorporate larger interactions we can sacrifice ease of computation for a larger Hilbert space. A weaker restriction is to consider the four bottom layers of the Hilbert space tree ($N_2=0,1,2,3$). The basis contains $4N-2$ states, so it still scales linearly with $N$, but the Hamiltonian is not tridiagonal. The benefit is being able to investigate couplings to higher states of the Hilbert space in the presence of strong interactions. Figs.~\ref{fig:5ParticleParameterSpace}(e,f) show the calculations for the four-layer restricted system. We immediately see that the inclusion of basis states with $N_2=2,3$ is necessary to show that CTAP is unstable in the high interaction, low detuning regime, as shown, for example, in Fig.~\ref{fig:5ParticleHilbertTrajectories}(b). This concept of expanding the Hilbert space from the bottom up, rather than considering the whole space, is analogous to the way matrix product states minimize entanglement by exploring the Hilbert space as required \cite{MPS}. To determine the point at which this instability occurs we need to consider the energies of the basis states to find limits in the parameter space for when CTAP works.



\subsection{Limiting the CTAP parameter space}
All the models show that CTAP is not possible in the high detuning, high interaction regime. Although we have seen that detuning can recover CTAP in the presence of interactions, too much detuning will still break the adiabaticity of the transport. The solid curves in Figs.~\ref{fig:5ParticleParameterSpace}(a,c,e) are contours of constant adiabaticity, calculated from Eq.~\eqref{eq:adiabatic_int}, for ${\cal A}_{1,0-}=0.05, 0.06, 0.07$. For a given finite pulsing time $\tau$ these curves accurately predict the limit in the parameter space where adiabatic transport is possible. Increasing $\tau$ will push this boundary further out and allow stronger interactions relative to the detuning of well 2.

\begin{figure}[t!]
\centering
\includegraphics[]{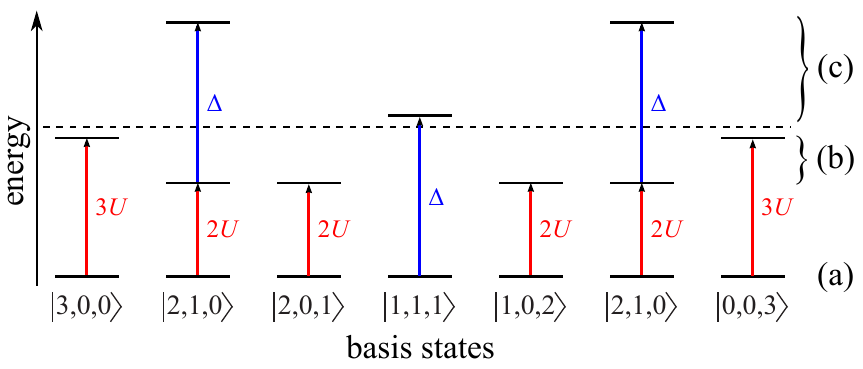}
\caption{Relative energies of basis states in the $N_2=0$ and $N_2=1$ subspaces for a system of three particles. (a) With no interactions ($U=0$) and no detuning ($\Delta=0$) these states are all degenerate. (b) In the interacting system with no detuning ($\Delta=0$), the interaction energy ($\sum_i U N_i (N_i-1)/2$) shifts each state differently (red shifts). Note that this includes $|1,1,1\rangle$, which has zero interaction energy. (c) Detuning the middle well in the interacting system shifts the $N_2=1$ states by the same amount (blue arrows). The dashed line indicates that there exists a minimum detuning for which all $N_2=1$ states will have higher energy than all $N_2=0$ states.}
\label{fig:BasisEnergies}
\end{figure}

The other region of interest is the low detuning, high interaction regime, $\Delta \lesssim NU$. In this regime the two-layer Hilbert space calculation, shown in Figs.~\ref{fig:5ParticleParameterSpace}(c,d), significantly deviates from the full and four-layer Hilbert space calculations, shown in Figs.~\ref{fig:5ParticleParameterSpace}(a,b) and (e,f), respectively. In analogy to the non-interacting case in Fig.~\ref{fig:Spectrum}(c), this is where the detuning is low enough for some eigenstates to cross the interacting CTAP state; the adiabaticity of the transport is broken by these crossing states in the small detuning regime, and not necessarily at $t=\tau/2$, which is typically the time when the adiabaticity parameter is maximized. To understand the breakdown of CTAP in this region we need to consider the energies of the basis states in the presence of interactions. Empirically, we find that efficient CTAP only occurs when the maximum energy of the $N_2=0$ states of the Hilbert space is less than the minimum energy of the $N_2=1$ states of the Hilbert space, as shown in Fig.~\ref{fig:BasisEnergies}. That is, there is a minimum detuning which makes $N_2=1$ states -- which couple to higher states -- energetically unfavourable, therefore resisting any attempt by the system to evolve into higher Hilbert space layers. For $N$ particles the highest energy basis state of the $N_2=0$ layer of the Hilbert space is 
\begin{alignat}{1}
E_{|N,0,0\rangle}=&\frac{U}{2}\left(N^2-N\right)
\end{alignat}
and the lowest energy basis state of the $N_2=1$ layer of the Hilbert space is
\begin{alignat}{2}
E_{| \frac{N-1}{2},1,\frac{N-1}{2}\rangle} &= 2\times \frac{U}{2}\left(\frac{N-1}{2}\right)\left(\frac{N-3}{2}\right) + \Delta, \\
E_{|\frac{N-2}{2},1,\frac{N}{2}\rangle} &=  E_{|\frac{N}{2},1,\frac{N-2}{2}\rangle} \nonumber \\
&= \frac{U}{2}\left(\frac{N-2}{2}\right)\left(\frac{N-4}{2}\right) \nonumber \\ 
& \quad +\frac{U}{2}\left(\frac{N}{2}\right)\left(\frac{N-2}{2}\right) + \Delta,
\end{alignat}
for an odd and even number of particles respectively. Hence, the criterion that the highest energy of the $N_2=0$ states is less than the lowest energy of the $N_2=1$ states is given by
\begin{equation}
\Delta > 
\begin{cases}
NU \left(\frac{N}{4}+\frac{1}{2}-\frac{3}{4N}\right) & \text{odd } N, \\
NU \left(\frac{N}{4}+\frac{1}{2}-\frac{3}{4N}\right) & \text{even } N.
\end{cases}
\label{eq:Delta_odd}
\end{equation}
These inequalities give the minimum detuning needed to efficiently transport $N$ interacting particles across the three wells. The region to the right of the dashed line plotted in Figs.~\ref{fig:5ParticleParameterSpace}(a,c,e) satisfies Eq.~\eqref{eq:Delta_odd} for $N=5$, while transport in the region to the left is unstable. 

\subsection{Comparison with Gross-Pitaevskii analysis}
We finally turn our attention to previous studies of CTAP in dilute gas BECs \cite{Graefe,Wang,Liu,Rab,Nesterenko}, which have focused on a Gross-Pitaevskii mean-field three-mode model. Following \cite{Rab} we consider a one dimensional system, with each of the three wells containing a single mode, equivalent to the Bose-Hubbard system schematically shown in Fig.~\ref{fig:System}. 
The energy of each mode is the ground state energy of each well, $U$ is again the interaction strength (equivalent to $g$ in \cite{Rab}), $\Delta$ is the detuning of the second mode relative to the left and right modes, and $K_{ij}$ describe the wavefunction overlap, and hence tunneling rate between modes $i$ and $j$. In the Gross-Pitaevskii analysis the quantities of interest are the populations $N_i(t)$ of mode $i$. Using the same squared-sinusoidal pulsing of Eq.~\eqref{eq:pulses}, Figs.~\ref{fig:5ParticleParameterSpace}(g,h) plot the final population of mode three, $N_3(t=\tau)/N$, and the maximum population of well two, $\max N_2(t)/N$, respectively. The regimes of high fidelity CTAP in the three-mode Gross-Pitaevskii analysis qualitatively agree with the Bose-Hubbard models. The two models are typically applied to very different limits, the former being a semi-classical mean-field model for a large number of particles, and the latter a linear quantum model that allows for entanglement. The agreement of the two models demonstrates the robustness of the CTAP protocol.


\section{CONCLUSIONS}
We have considered adiabatic transport of $N$ particles across a three-well Bose-Hubbard system and solved the non-interacting case analytically for which we obtained the null space and energy eigenspectrum in closed form. Detuning well 2 relative to wells 1 and 3 makes the null state unique and it is this state in which $N$ particles can be adiabatically transported from well 1 to 3, or vice-versa, without occupation of well 2. Even without detuning, CTAP is possible if the system is initialized correctly. The non-interacting $N$-particle three-well system exhibits similar dynamics to adiabatic transport of single particles along longer chains of $2N+1$ sites. 

The interacting Bose-Hubbard three-well system cannot be solved analytically and there is no general null state but the time-dependent Schr\"odinger equation can be integrated numerically for few particles. We have shown that for stronger interactions relative to the middle well detuning, which disrupt the transport, CTAP can be maintained by detuning well 2 enough to isolate the CTAP state from other energy eigenstates, but not so much as to break the adiabaticity of the transport. 

By considering the weakly interacting case as a perturbation of the non-interacting case we have defined useful limits for when adiabatic transport is possible using a squared-sinusoidal pulsing. These limits stem from minimizing the adiabaticity parameters, i.e.~the coupling between the CTAP state and the closest eigenstate, and give good agreement to the limit of detuning and interaction strength for which adiabatic transport is possible for a finite pulsing time.

Having established the detuning-interaction regime for which CTAP is possible we have investigated restricting the Hilbert space to the minimum number of basis states necessary to allow CTAP. For zero or very small interactions relative to the detuning only states with $N_2=0$ or $1$ need to be included in the basis. For interaction energies comparable to or greater than the detuning, we must include states with $N_2>1$, but we can still restrict the Hilbert space to the lowest layers to allow CTAP. Apart from the adiabaticity criterion Eq.~\eqref{eq:adiabatic_int}, which holds for the restricted and unrestricted Hilbert spaces, we have obtained a limit on what defines this strongly interacting regime by comparing the energy of individual basis states. In this regime we have seen that adiabatic transport is not stable.

Finally, we have compared CTAP in the quantum Bose-Hubbard model to CTAP in a three-mode mean-field Gross-Pitaevskii approximation. The regime of high fidelity of CTAP in both models matches and clearly demonstrates the robustness of the CTAP protocol in both the quantum and semi-classical mean-field limits. This mapping between the mean-field approach and the Bose-Hubbard model reinforces the use of the semi-analytic constraints we have developed to determine the validity of CTAP.  

\begin{acknowledgments}
A.D.G. acknowledges the Australian Research Council for financial support (Project No. DP0880466).
\end{acknowledgments}

\end{document}